\documentclass[twoside,12pt]{article}
\usepackage{epsfig}

\newcommand{\be}{\begin{equation}}
\newcommand{\ee}{\end{equation}}
\newcommand{\bea}{\begin{eqnarray}}
\newcommand{\eea}{\end{eqnarray}}

\begin{document}
\title{ \vspace{1cm} Particle asymmetries in the early universe}
\author{Maik Stuke \\
Fakult\"at f\"ur Physik, Universit\"at Bielefeld, Postfach 100131, 33501 Bielefeld, Germany.}
\maketitle
\begin{abstract}
The total lepton asymmetry $l=\sum_f l_f$ in our universe is only poorly constrained by theories and experiments. It might be orders of magnitudes larger than the observed baryon asymmetry $b\simeq {\cal O}(10^{-10})$, $|l|/b \leq {\cal O}(10^{9})$. We found that the dynamics of the cosmic QCD transition changes for large asymmetries. Predictions for asymmetries in a single flavour $l_f$ allow even larger values. We find that asymmetries of $l_f\leq {\cal O}(1)$ in a single or two flavours change the relic abundance of WIMPs. However, large lepton and large individual lepton flavour asymmetries influence significantly the dynamics of the early universe. 
\end{abstract}
%
%
%
\section{Introduction}
%
In this article we want to describe the thermodynamic evolution of the standard model particles in the early universe. 
We will focus on two main events after the electroweak transition at $T_{ew}\simeq 200$ GeV \cite{Laine:2000xu}: the freeze out of WIMPs between 40 GeV and 400 MeV \cite{Green:2005fa} and in more detail the cosmic QCD transition at $T_{QCD}\simeq 200$ MeV \cite{Schwarz:2009ii}.
  
All standard model particles can be described as a particle fluid in chemical equilibrium. This excellent approximation holds because all interaction rates are much larger then the Hubble rate $H$. Within the Hubble time $t_H=1/H$ particles get created and destroyed so fast, that the chemical equilibrium is always fulfilled for $T>$ few MeV. All particle interactions conserve charge-, baryon- and lepton flavour number. Since our universe is charge neutral \cite{Siegel:2006px}, the sum of all charged particles has to be zero. We know further from big bang nucleosythesis or the WMAP measurements, that in our universe are slightly more baryons then anti baryons. The difference in their densities normalized with the entropy density $s(T)$ is measured as $b= {\cal O}(10^{-10})$ \cite{Komatsu:2008hk}. Things are more difficult for the lepton flavour number. In the temperature region of our interest, 200 GeV $>T>$ few MeV, each lepton flavour is conserved. Neutrino oscillations would violate this assumption, but they become sufficient at much lower temperatures $T<T_{osc}\simeq10 MeV$. The lepton flavour density is defined as $l_f= (n_f + n_{\nu_f})/s(T)$ for $f=e$, $\mu$, $\tau$. $n_f$ is the net number density of the charged lepton flavour and $n_{\nu_f}$ the one of the neutrinos. The latter one is unknown and it might be, that the asymmetry between neutrinos and anti neutrinos is orders of magnitude larger then $b$. From big bang nucleosythesis, the  cosmic microwave background and the large scale structure we only know $|l_f|\leq 0.2$ for $T<T_{osc}$ \cite{Simha:2008mt}. This assumes the asymmetry to be equal in all flavours $l_e=l_{\mu}=l_{\tau}$. For $T>T_{osc}$ this does not necessarily hold. It was shown that initial different flavour asymmetries equilibrate due to the neutrino oscillations \cite{Dolgov:2002ab} before big bang nucleosynthesis. A scenario with asymmetries for example like $l_e=b\simeq10^{-10}$ but $l_{\mu}=-l_{\tau}={\cal O}(1)$ would be possible \cite{Casas:1997gx,MarchRussell:1999ig} and in no contradiction to experimental data \cite{Simha:2008mt}.  

Leaving the lepton flavour asymmetries as free parameters, we can set up the following system of 5 equations for the evolution of the standard model particles in the early universe:
\begin{eqnarray}
\label{lf} l_f&=&\frac{n_f+n_{\nu_f}}{s(T)}\\
\label{bary} b&=&\sum_i \frac{b_in_i}{s(T)} \\
\label{charge} 0&=&\sum_i q_in_i
\end{eqnarray}       
with $f=e$, $\mu$ or $\tau$, the baryon number $b_i$ of species $i$, and $q_i=$ the charge of species $i$. In a Friedman-Lemaitre-Robertson-Walker universe the net number density of a particle $i$ and its antiparticle $\bar{i}$ with chemical potential $\mu_i=-\mu_{\bar{i}}$ at a temperature $T$ is given by \cite{Schwarz:2009ii}
\begin{eqnarray}
n_i(T,\mu_i)&=&\frac{g_i}{2\pi^2}\int^{\infty}_{m_i}{E(E^2-m^2)^{1/2}\left(f(i)-f(\bar{i}) \right)\rm{d}E}\nonumber \\
\label{ni}&=&\frac{g_i}{2\pi^2}\int^{\infty}_{m_i}{E\sqrt{E^2-{m_i}^2}\left(\frac{1}{\rm{exp}\frac{E-\mu_i}{T}\pm1}-\frac{1}{\rm{exp}\frac{E+\mu_i}{T}\pm1} \right)}dE,  
\end{eqnarray}
where $g_i$ counts the degrees of freedom of the particle and the integral over the energy $E$ runs from the mass $m_i$ to infinity. The $\pm$ depends on the statistics: $-$ for bosons and $+$ for fermions. 
For fermions this can be approached to
\begin{eqnarray}
\label{nirel}n_i(T,\mu_i)&\stackrel{T\gg m_i,\mu_i}{=}& \frac{1}{6}g_i T^3 \frac{\mu_i}{T}+{\cal O}\left(\frac{\mu_i}{T}\right)^3,\\
\label{nimass}&\stackrel{T\ll m_i}{=}&2g_i\left(\frac{mT}{2\pi} \right)^{3/2} \rm{sinh}\left(\frac{\mu_i}{T}\right)\rm{exp}\left(\frac{m_i}{T}\right).
\end{eqnarray}
The chemical potentials of all particles in chemical equilibrium can be expressed by five independent, e.g. $\mu_{\nu_e},\mu_{\nu_{\mu}},\mu_{\nu_{\tau}}, \mu_e,$ and $\mu_u$. For details see \cite{Schwarz:2009ii}. For a given temperature we solved (\ref{lf}) to (\ref{charge}) for any lepton flavour asymmetry. With the calculated chemical potentials we calculated the energy density of each particle, given by
\begin{equation}
\epsilon_i=\frac{g_i}{2\pi^2}\int^{\infty}_{m_i}{E^2\sqrt{E^2-{m_i}^2}\left(\frac{1}{\rm{exp}\frac{E-\mu_i}{T}\pm1}\right)}dE. 
\end{equation}
If the cooling universe reaches $T\simeq (1/3)m_i $ the energy becomes too low to create the particle $i$ and it  annihilates. The effective degrees of freedom $g_{\ast}$ of the particle fluid gets reduces by the number of effective degrees of freedom of the annihilating particle. The definition of the $g_{\ast}$ is 
\begin{equation}
\label{gast}g_{\ast} = \frac{30}{\pi^2T^4}\sum_j \epsilon_j.
\end{equation}
A more detailed discussion can be found for instance in \cite{Schwarz:2009ii,Schwarz:2003du}.
In the following chapter we present our results for temperatures around the QCD transition for the case of equal distributed flavour asymmetries $l=\sum_f l_f=3l_f$. Another interesting scenario is the influence of $b=l$ with $l_{\mu}=-l_{\tau}\gg l_{e}\simeq b$ on the degrees of freedom, which is presented in chapter 3. We conclude our results in chapter 4.

%
%
%
\section{The cosmic QCD transition}
\begin{figure}[h]
	\centering
		\includegraphics[width=0.30\textwidth, angle=270]{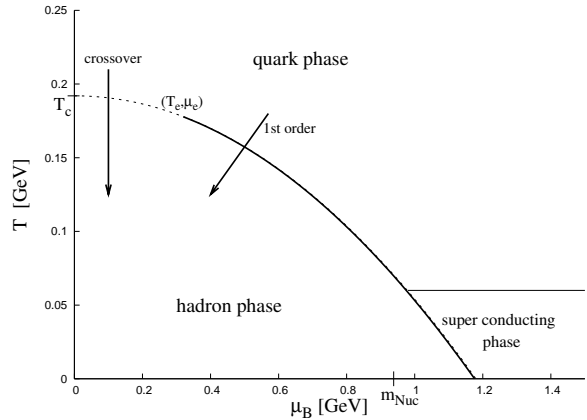}
	\caption{QCD diagram inspired by perturbative calculations, heavy ion collisions, and lattice simulations. If the trajectory crosses the dotted line, the transition is a crossover, whereas crossing the solid line would indicate a first order transition. The solid line may end in a critical point at $(T_e,\mu_e$). The temperature of the transition is $T_{c}\simeq 200$ MeV \cite{Stephanov:2007fk,Philipsen:2008gf}. The trajectory of the universe has to start at high temperatures in the quark phase and end at $\mu_b(T=0)=m_{Nuc}$, the nucleon mass in the hadron phase.}
	\label{fig:figure1}
\end{figure}

%
The standard model of particle physics predicts a spontaneous breaking of chiral symmetry of the QCD. This leads to a phase transition at a certain temperature $T_{QCD}$ where quarks confine to hadrons. One would like to have a phase diagram to locate the different states of matter. Such a phase diagram would than be spanned by the temperature and the baryo chemical potential $\mu_B$. Unfortunately the calculations for such a diagram from first principles are extraordinary difficult \cite{Stephanov:2007fk,Philipsen:2008gf}. So far a combination of results from perturbative calculations, heavy ion collisions and lattice simulations lead to the diagram in figure \ref{fig:figure1}.   

For the evolution of the early universe the cosmic QCD transition is of crucial interest for our understanding of the early universe. One would like to have more knowledge about how quarks confine to hadrons, since it sets for example the starting conditions for one of the pillars of cosmology, the big bang nucleosynthesis. Depending on the order of the transition there might be several today observable relics like quark nuggets, magnetic fields, and the modification of the primordial background of gravitational waves \cite{Schwarz:2003du,Caprini:2010xv}. If the transition was of first order, the universe might have undergone a second period of inflation, suggested by \cite{Boeckel:2009ej}. We want to review here our results on the effect of leptons on the transition \cite{Schwarz:2009ii} and show, how they change figure \ref{fig:figure1}.

Each of the five conserved quantum numbers $b$, $l_f$, and $q$ in equations (\ref{lf}) to (\ref{charge}) can be associated with a chemical potential, $\mu_B$, $\mu_Q$, and $\mu_{L_f}$, which can be expressed by the chemical potentials of the standard model particles \cite{Schwarz:2009ii}. 
To get an estimate of the $l$ and $b$ dependence of the chemical potentials we assume for $T\gg T_{QCD}$ up, down, charm and strange quarks and the leptons massless. Solving equations (\ref{lf}) to (\ref{charge}) with (\ref{nirel}) leads to
\begin{eqnarray}
\label{muB}\mu_B(T\gg T_{QCD})&=& \left(\frac{39}{4}b - l \right)\frac{s(T)}{4T^2}\\
\label{muQ}\mu_Q(T\gg T_{QCD})&=& \left(-\frac{3}{4}b + l \right)\frac{s(T)}{2T^2}\\
\label{muLf}\mu_{L_f}(T\gg T_{QCD})&=&\left(-\frac{1}{4}b + l \right)\frac{s(T)}{T^2}.
\end{eqnarray}
If the total lepton asymmetry $l$ is orders of magnitudes larger, it becomes the dominant factor for all three chemical potentials. The order of $\mu_B$ is then given by the order of $l$. We see also, that the chemical potentials can get negative, depending on the sign of $l$. 
For temperatures below pion mass shell we get for the baryo chemical potential the following estimates:
\begin{equation}
\label{muBnr}\mu_B(T<(m_{\pi}/3T))\approx m-T\rm{ln}\left[\frac{c(T)}{2bs(T)}\right],
\end{equation}
with $c(T)=4(\frac{Tm}{2\pi})^{3/2}$. Independent of $l$ all trajectories for the baryo chemical potential are unique for $T<(m_{\pi}/3T)\simeq50$ MeV. 
All effects are shown in our full numerical calculations in figures \ref{fig:mub} and \ref{fig:muLe}. For these pictures we calculated (\ref{lf}) to (\ref{charge}) with net particle densities (\ref{ni}) for all standard model particles including their physical masses. 
%
%
%
\section{Effective degrees of freedom and WIMPs}
%
In this chapter we want to give a brief introduction to the influence of large individual lepton flavour asymmetries on the relic abundance of WIMPs. A more detailed discussion can be found in \cite{tbp}.

In the following we assume the dark matter to be a single component WIMP with no asymmetry between WIMPs and anti WIMPs. These particles with masses $m_{\chi}$ typically between 10 GeV and 1 TeV decouple chemically from the standard model particles between 40 GeV $>T_f>0.4$ GeV \cite{Green:2005fa} depending on their mass. After this freeze out the WIMPs only react kinetically with the standard model particles. Their relic abundance Y can be approximated by \cite{Green:2005fa}
\begin{equation}\label{relAb}
\rm{Y}\simeq \left(\frac{45}{\pi} \right)^{1/2} \frac{1}{m_{\chi}M_p\left\langle \sigma |v| \right\rangle_{T_{fo}}} \frac{x_{fo}}{\sqrt{{g}_{\ast}}}.
\end{equation}
$M_P$ is the Planck mass, $\sigma$ the total cross section and $v$ the velocity of the annihilating particles. The index $fo$ indicates the freeze out time and $x_{fo}=m_{\chi}/T_{fo}\propto \rm{ln}(1/g_{\ast})$. 

The relic abundance depends on the relativistic degrees of freedom $g_{\ast}$ of the standard model particle ensemble. Any few percent effect (\ref{gast}) would lead to a few percent effect in the relic abundance of the WIMPs. 
To get an analytic approximation of of the $l_f$ dependence of (\ref{gast}) we take all particles massless and get
\begin{eqnarray}
g_{\ast}(T,\{\mu_i\}) &\stackrel{m_i=0}{=}& \frac{15}{T^4\pi^4}\sum_i g_i \int_0^{\infty}{\frac{E^3}{\rm{exp}\left[\frac{E-\mu}{T} \right]+1}{\rm d}E}\\
&=&\sum_F \frac{7}{4}g_F + \frac{15}{2}g_F\left(\frac{\mu_F}{\pi T} \right)^2 + \frac{15}{4}g_F\left(\frac{\mu_F}{\pi T} \right)^4 + \sum_B g_B.
\end{eqnarray} 
The index $F$ denotes fermions and $B$ bosons. Compared to the standard case where chemical potentials are neglected, we have additional 
\begin{equation}
\Delta g_{\ast}= \frac{15}{2}\sum_F{g_i\left(\left(\frac{\mu_i}{\pi T}\right)^2 + \frac{1}{2}\left(\frac{\mu_i}{\pi T}\right)^4\right)}.
\end{equation}
If the fermion chemical potentials $\mu_F/T$ are ${\cal O}(10^{-1})$, the effect becomes non negligible. If for example the asymmetry in the $\mu$- and $\tau$ neutrinos is of order unity, their chemical potentials would become that big.  
We calculated (\ref{gast}) numerically for the full standard model and for a scenario where $\sum_fl_f=b$ with $l_e=b$ and $l_{\mu}=l_{\tau}$. If these asymmetries are in the neutrinos, values of $\mu_{\nu_{\mu}}=\mu_{\nu_{\tau}}={\cal O}(0.5)$ are not in contradiction to theories \cite{McDonald:1999in,Campbell:1998yi,MarchRussell:1999ig,Casas:1997gx} nor to experiments \cite{Dolgov:2002ab,Simha:2008mt}. Our numerical calculations of $\ref{gast}$ for the whole set of standard model particles show a significant effect.
\begin{figure}[h]
	\centering
		\includegraphics[width=0.30\textwidth, angle=270]{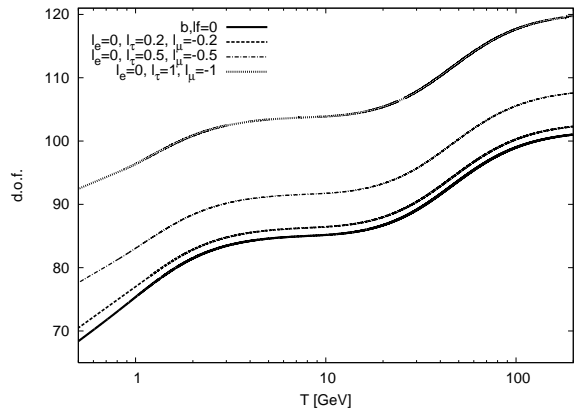}
	\caption{The degrees of freedom calculated numerically for the scenario $l_e=0$ and $l_{\mu}=-l_{\tau}$. For $l_{\mu}\leq 0.1$ there is a significant deviation compared to the $l_i=0$ scenario. For $l_{\mu}=1$ we observe a more then 20$\%$ effect.}
	\label{fig:dof}
\end{figure}
%
%
%
\section{Conclusion}
%
Our knowledge of the relic neutrinos and the lepton asymmetries is still very poor and the possible effects of large lepton (flavour) asymmetries on the early universe have been overlooked so far. We investigated here the effect of large $l$ and $l_f$ for $T_{ew}>T>T_{\nu -osc}$ and found that large lepton (flavour) asymmetries do have an observable impact on the dynamics of the early universe. 

An asymmetry $b\ll|l|\leq 0.01$ influences significantly the dynamics of the QCD transition \cite{Schwarz:2009ii}. Depending on the unknown phase structure of the QCD phase diagram in the $\mu_B-T$ plane and the position of the hypothetical critical end point, a large lepton asymmetry might result in a first order transition. This seems to be possible for $l\geq 0.02$.

Large lepton asymmetries determine the order of the baryo chemical potential. In general, $\mu_B$ is a function of both, the baryon and lepton asymmetry, $\mu_B=\mu_B(T,l,b)$. 

However, we have shown that the phase diagram for the cosmic QCD phase transition needs at least six dimensions, one for each chemical potential and the temperature. The structure of these phase diagrams needs further investigations, for instance the inclusion of interactions, to clarify the order.  

We have also introduced in this article the idea, that the unknown neutrino flavour asymmetries $l_f\geq 1$ effect the relic abundance of the WIMP dark matter. Even if the total lepton asymmetry is of the same order as the baryon asymmetry $l=b$, but individual flavour asymmetries  $l_i\leq{\cal O}(1)$, we observed a few percent effect.

In our discussion of this work we did not include sterile neutrinos and in particular not their oscillations with active neutrinos. The inclusion would possibly weaken the constraints on the neutrino asymmetries in the early universe.
%
%
%
\section{Acknowledgment}
I would like to thank the organizers for this inspiring workshop and for the possibility to present this talk. My stay at the conference was supported by a DFG fellowship. I am grateful to Dominik J. Schwarz, with whom this work was done. My work is founded by the Friedrich-Ebert-Foundation.
%
%
%

%
%
%
\begin{figure}[htb]
	\centering
		\includegraphics[width=0.30\textwidth, angle=270]{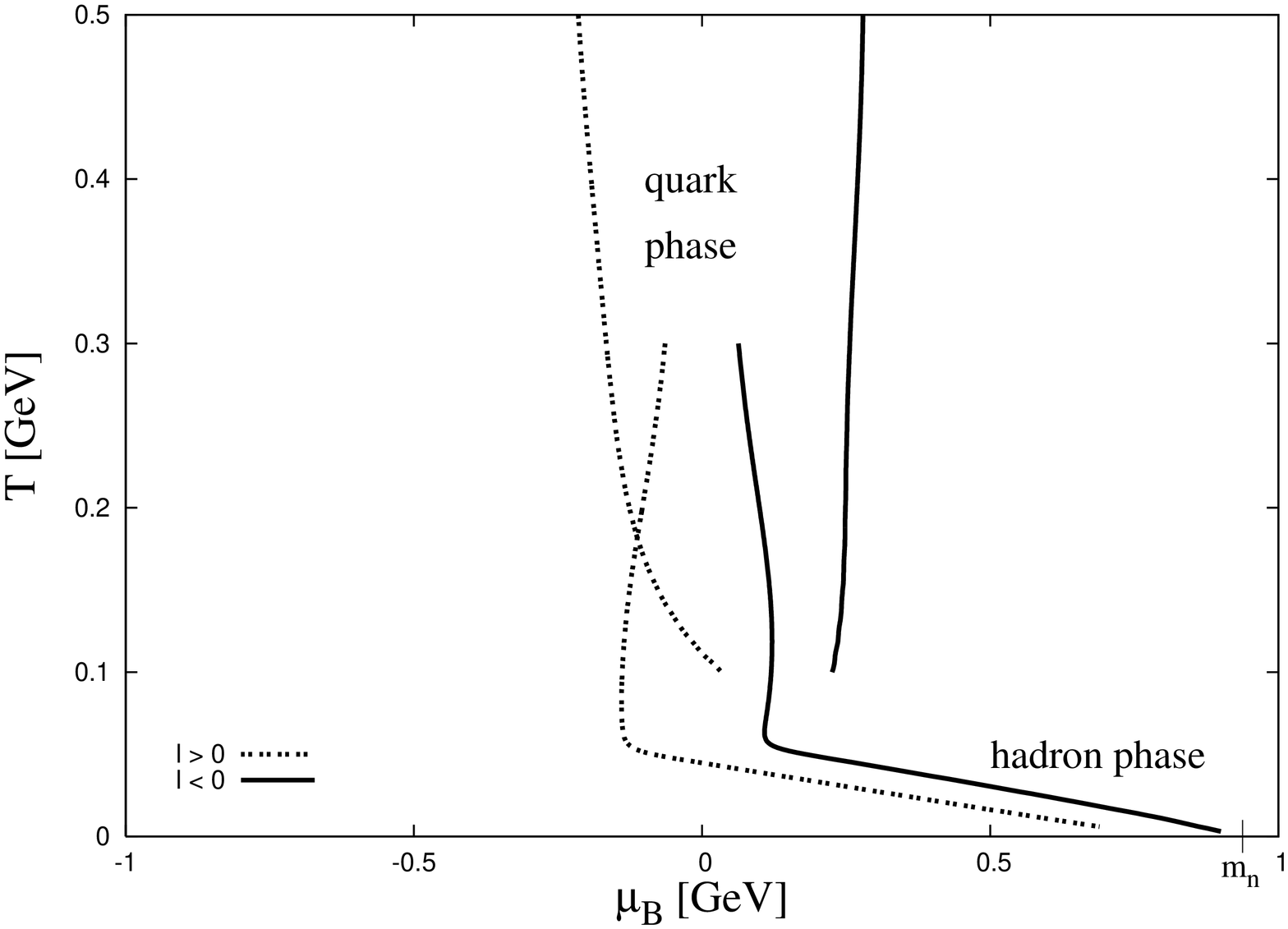}
		\hspace{1mm}
		\includegraphics[width=0.30\textwidth, angle=270]{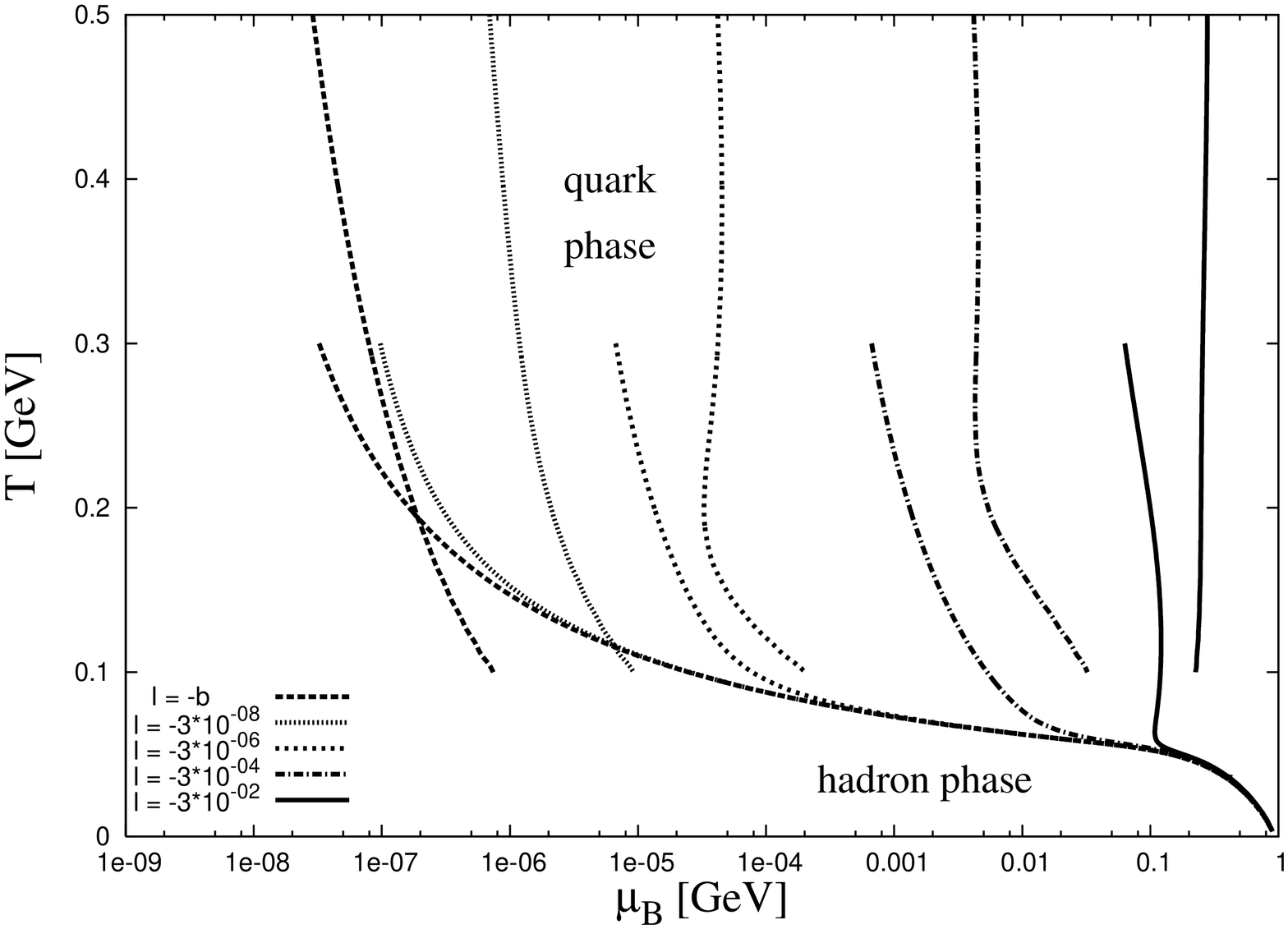}
	\caption{Phase diagram for the baryo chemical potential. On the left hand is the sign dependence of $\mu_B$ shown. In the high temperature regime a large and positive $l$ would result in a negative trajectory, while a negative $l$ leads always to a positive $\mu_B$, as indicated by (\ref{muB}). In the low temperature regime of the hadron phase, both cases approach the nucleon mass. On the right hand picture are the trajectories for different $l$ asymmetries shown.}
	\label{fig:mub}
	\vspace{1mm}

	\centering
		\includegraphics[width=0.30\textwidth, angle=270]{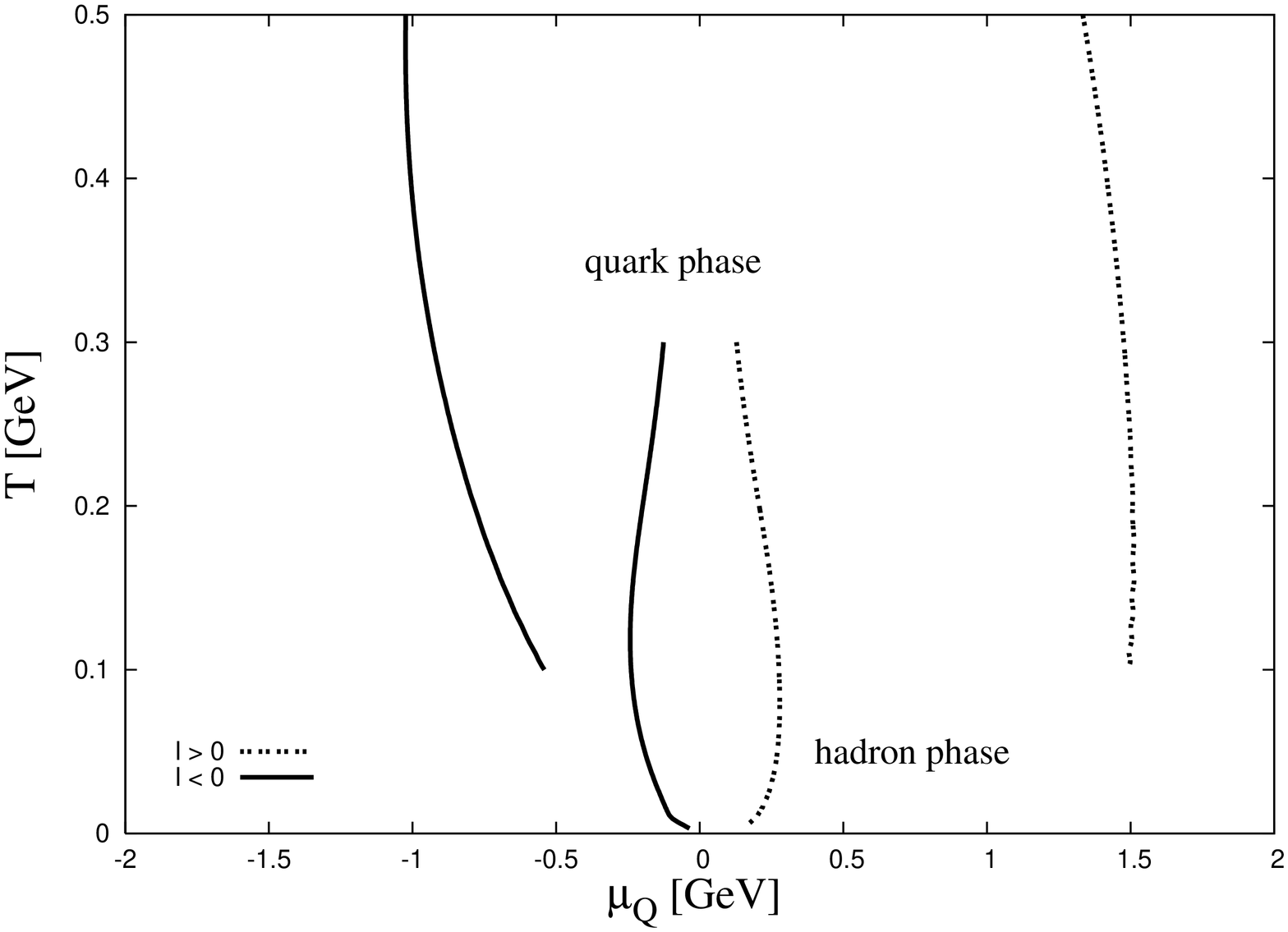}
		\hspace{1mm}
		\includegraphics[width=0.30\textwidth, angle=270]{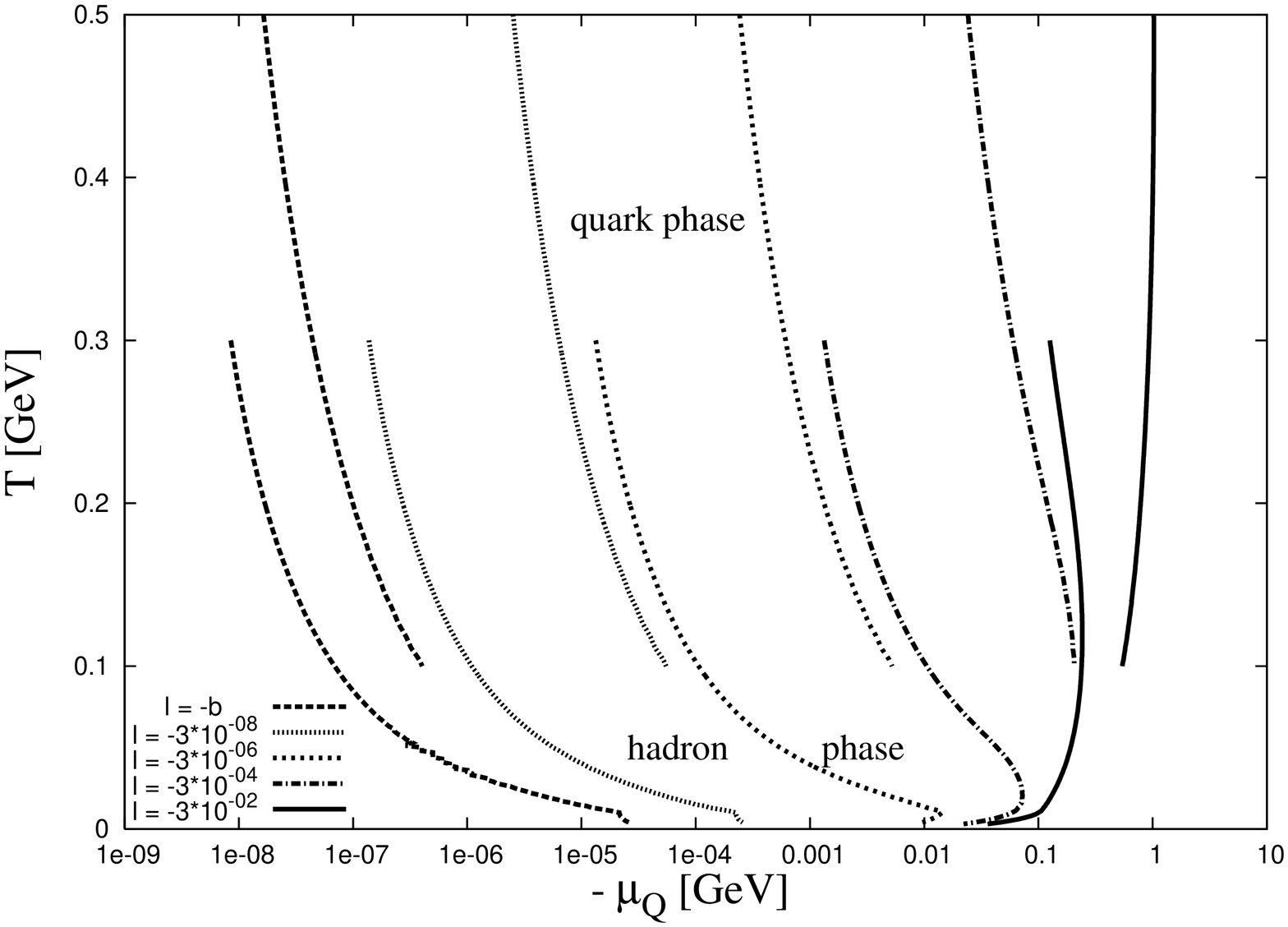}
		\vspace{1mm}

		\includegraphics[width=0.30\textwidth, angle=270]{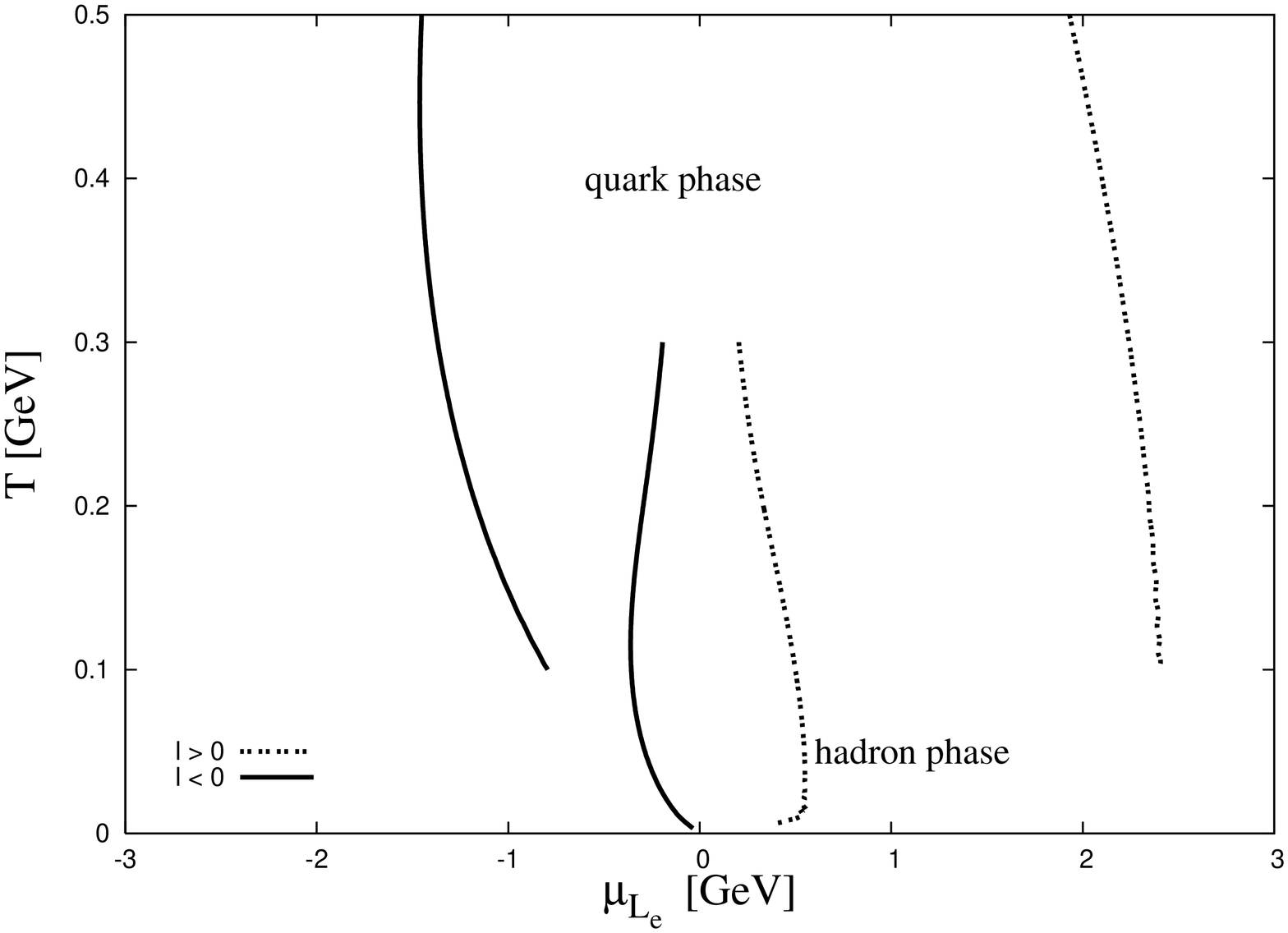}
		\hspace{1mm}
		\includegraphics[width=0.30\textwidth, angle=270]{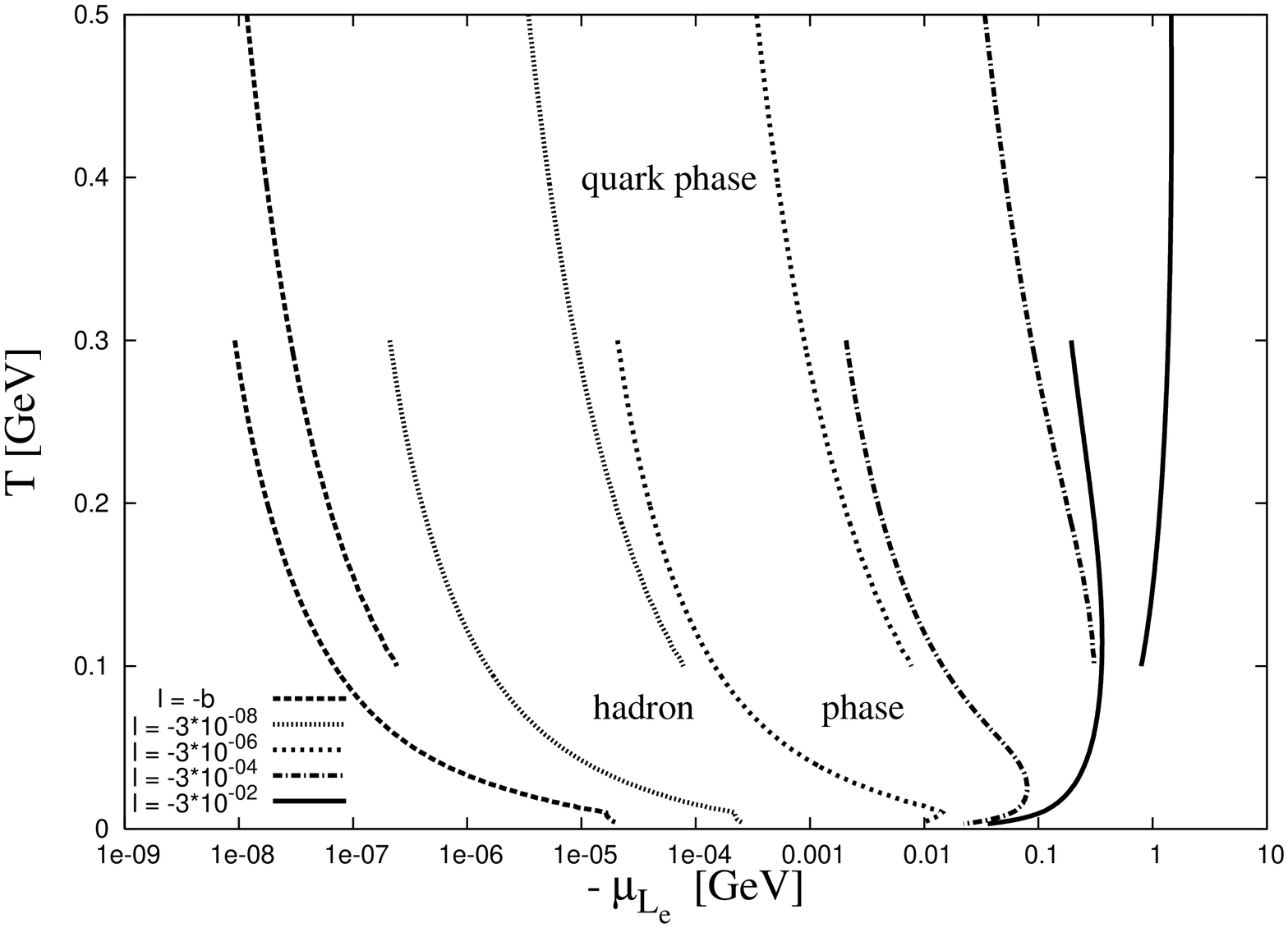}
	\caption{phase diagrams for the charge and electron type lepton flavour chemical potential. The upper two pictures show the trajectories of the charge chemical potential and the lower the trajectories of the electron flavour chemical potential. Both chemical potentials depend on the same way on the sign of large $l$. Also here the splittings of the trajectories for different $l$ are shown.}
	\label{fig:muLe}
\end{figure}


\begin{thebibliography}{99}
\itemsep -2pt 

\bibitem{Schwarz:2009ii}
  D.~J.~Schwarz and M.~Stuke,
  JCAP {\bf 0911} (2009) 025
  [arXiv:0906.3434 [hep-ph]].

\bibitem{Schwarz:2003du}
  D.~J.~Schwarz,
  Annalen Phys.\  {\bf 12} (2003) 220
  [arXiv:astro-ph/0303574].

\bibitem{Siegel:2006px}
  E.~R.~Siegel, J.~N.~Fry,
  Astrophys.\ J.\  {\bf 651 } (2006)  627-635.
  [astro-ph/0604526];
E.~R.~Siegel, J.~N.~Fry,
    [astro-ph/0609031].

\bibitem{Komatsu:2008hk}
  E.~Komatsu {\it et al.}  [WMAP Collaboration],
  Astrophys.\ J.\ Suppl.\  {\bf 180} (2009) 330
  [arXiv:0803.0547 [astro-ph]].

\bibitem{Simha:2008mt}
  V.~Simha and G.~Steigman,
  ``Constraining The Universal Lepton Asymmetry,''
  JCAP {\bf 0808}, 011 (2008)
  [arXiv:0806.0179 [hep-ph]].

  J.~Dunkley {\it et al.},
  arXiv:1009.0866 [astro-ph.CO].

  L.~M.~Krauss, C.~Lunardini, C.~Smith,
    [arXiv:1009.4666 [hep-ph]].

  L.~A.~Popa and A.~Vasile,
  JCAP {\bf 0806}, 028 (2008)
  [arXiv:0804.2971 [astro-ph]].
  S.~Pastor, T.~Pinto and G.~G.~Raffelt,
  Phys.\ Rev.\ Lett.\  {\bf 102}, 241302 (2009)
  [arXiv:0808.3137 [astro-ph]].

\bibitem{Laine:2000xu}
  M.~Laine,
  arXiv:hep-ph/0010275, and references therein.

\bibitem{Dolgov:2002ab}
  A.~D.~Dolgov, S.~H.~Hansen, S.~Pastor {\it et al.},
  Nucl.\ Phys.\  {\bf B632 } (2002)  363-382.
  [hep-ph/0201287];
  Y.~Y.~Y.~Wong,
  Phys.\ Rev.\  {\bf D66}, 025015 (2002).
  [hep-ph/0203180].


\bibitem{McDonald:1999in}
  J.~McDonald,
  Phys.\ Rev.\ Lett.\  {\bf 84}, 4798 (2000)
  [arXiv:hep-ph/9908300].

\bibitem{Campbell:1998yi}
  B.~A.~Campbell, M.~K.~Gaillard, H.~Murayama and K.~A.~Olive,
  Nucl.\ Phys.\  B {\bf 538} (1999) 351
  [arXiv:hep-ph/9805300].


\bibitem{MarchRussell:1999ig}
  J.~March-Russell, H.~Murayama and A.~Riotto,
  JHEP {\bf 9911}, 015 (1999)
  [arXiv:hep-ph/9908396].


\bibitem{Casas:1997gx}
  A.~Casas, W.~Y.~Cheng and G.~Gelmini,
  Nucl.\ Phys.\  B {\bf 538}, 297 (1999)
  [arXiv:hep-ph/9709289].

\bibitem{Stephanov:2007fk}
  M.~A.~Stephanov,
  PoS {\bf LAT2006 } (2006)  024.
  [hep-lat/0701002].

\bibitem{Philipsen:2008gf}
  O.~Philipsen,
  Prog.\ Theor.\ Phys.\ Suppl.\  {\bf 174 } (2008)  206-213.
  [arXiv:0808.0672 [hep-ph]].

\bibitem{Caprini:2010xv}
  C.~Caprini, R.~Durrer, X.~Siemens,
  Phys.\ Rev.\  {\bf D82 } (2010)  063511.
  [arXiv:1007.1218 [astro-ph.CO]];
  T.~Kahniashvili, L.~Kisslinger, T.~Stevens,
  Phys.\ Rev.\  {\bf D81 } (2010)  023004.
  [arXiv:0905.0643 [astro-ph.CO]].

\bibitem{Boeckel:2009ej}
  T.~Boeckel and J.~Schaffner-Bielich,
  Phys.\ Rev.\ Lett.\  {\bf 105} (2010) 041301
  [arXiv:0906.4520 [astro-ph.CO]].

\bibitem{tbp}
	G.~D.~Starkman, D.~J.~Schwarz, M.~Stuke, 
	``Relic WIMP abundance and lepton (flavour) asymmetries,''
	to be published.
	
\bibitem{Green:2005fa}
  A.~M.~Green, S.~Hofmann and D.~J.~Schwarz,
  JCAP {\bf 0508} (2005) 003
  [arXiv:astro-ph/0503387].

\bibitem{Jungman:1995df}
  G.~Jungman, M.~Kamionkowski and K.~Griest,
  Phys.\ Rept.\  {\bf 267} (1996) 195
  [arXiv:hep-ph/9506380].

\bibitem{Hindmarsh:2005ix}
  M.~Hindmarsh and O.~Philipsen,
  Phys.\ Rev.\  D {\bf 71} (2005) 087302
  [arXiv:hep-ph/0501232].

\end{thebibliography}
\end{document}